\documentclass[%
 aip,
 jmp,%
 amsmath,amssymb,
 reprint,%
]{revtex4-1}

\usepackage{graphicx}
\usepackage{dcolumn}
\usepackage{bm}

\begin{document}

\preprint{AIP}

\title[]{The influence of the secondary electron induced asymmetry on the Electrical Asymmetry Effect in capacitively coupled plasmas}

\author{Ihor Korolov}
\email{korolov.ihor@wigner.mta.hu}
\affiliation{Institute for Solid State Physics and Optics, Wigner Research Centre for Physics, Hungarian Academy of Sciences, P.O.B. 49, H-1525 Budapest, Hungary}

\author{Aranka Derzsi}
\affiliation{Institute for Solid State Physics and Optics, Wigner Research Centre for Physics, Hungarian Academy of Sciences, P.O.B. 49, H-1525 Budapest, Hungary}
\author{Zolt\'an Donk\'o}
\affiliation{Institute for Solid State Physics and Optics, Wigner Research Centre for Physics, Hungarian Academy of Sciences, P.O.B. 49, H-1525 Budapest, Hungary}
\author{Julian Schulze}
\affiliation{Institute for Plasma and Atomic Physics, Ruhr University Bochum, 44780 Bochum, Germany}

\date{\today}

\begin{abstract}
In geometrically symmetric capacitive radio-frequency plasmas driven by two consecutive harmonics a dc self-bias can be generated as a function of the phase shift between the driving frequencies via the Electrical Asymmetry Effect (EAE). Recently the Secondary Electron Asymmetry Effect (SEAE) was discovered (T. Lafleur, P. Chabert and  J.P. Booth  {\it J. Phys. D: Appl. Phys.} {\bf 46} 135201 (2013)): unequal secondary electron emission coefficients at both electrodes were found to induce an asymmetry in single frequency capacitive plasmas. Here, we investigate the simultaneous presence of both effects, i.e. a dual-frequency plasma driven by two consecutive harmonics with different electrode materials. We find, that the superposition of the EAE and the SEAE is non-linear, i.e. the asymmetries generated by each individual effect do not simply add up. The control ranges of the dc self-bias and the mean ion energy can be enlarged, if both effects are combined.
\end{abstract}

\maketitle

\noindent
Capacitively coupled radio frequency (CCRF) plasmas play a central role in modern plasma processing technologies \cite{LLBook}, for which the control of ion properties, viz. the ion flux, $\Gamma_{\rm i}$, and the mean ion energy, $\langle E_{i} \rangle$, is important. During the past years different strategies have been developed for the independent control of $\Gamma_{\rm i}$ and $\langle E_{i} \rangle$: (i) Discharges driven by two, or multiple (significantly different) frequencies \cite{Georgieva, Mussenbrock,Kawamura,Boyle,Kitajima} (ii) hybrid (inductive + capacitive) plasma sources \cite{Rauf_CI,Kawamura_rfdc_b,Schulze_Hybrid}, (iii) the use of customized voltage waveforms  \cite{Wendt,Patterson, Tailored1,Tailored4}, as well as (iv) discharges operated under the conditions of the Electrical Asymmetry Effect (EAE), where the driving voltage waveform contains the sum of multiple consecutive harmonics \cite{Heil, Heil_b, Donko1,EAE1exp}.

\noindent
Previous studies have shown that the EAE provides a better separate control of ion properties compared to ``classical'' dual-frequency discharges, where significantly different frequencies are used and the quality of this separate control is limited by frequency coupling effects as well as the presence of secondary electrons \cite{Donko_Gamma, Donko_Gammab}. Both are largely avoided by using the EAE. Considering the simplest case, i.e. two consecutive harmonics, the driving voltage (coupled via a capacitor to the discharge) for the generation of the EAE is:
\begin{equation}
\phi(t) = \phi_1 \cos (2\pi f t + \theta) + \phi_2 \cos (4 \pi f t).
\label{eq:EAEVolt}
\end{equation}

\noindent
The driving harmonics -- with fundamental frequency $f$ and amplitudes $\phi_1$ and $\phi_2$ -- are phase locked at an angle $\theta$. Using such a voltage waveform, a dc self-bias is generated as an almost linear function of $\theta$ for $0^\circ \le \theta \le 90^\circ$. The mean ion energy at the electrodes can be changed by a factor of about 2 at nearly constant $\Gamma_{\rm i}$ by tuning $\theta$ \cite{Heil,Donko1,EAE1exp}. Studies of the EAE have been performed for different gases \cite{EAE1exp,Chile,Wang,Longo,Hou,ourCF4}, harmonics' amplitudes \cite{EAEopt}, fundamental frequencies \cite{EAEFrVar}, and numbers of consecutive driving harmonics \cite{multi,EPS}.

\noindent
CCRF discharges can be operated in different electron heating modes such as the $\alpha$- and $\gamma$-mode \cite{BB}. While the former is dominant at low pressures and low voltage amplitudes as well as high driving frequencies, the $\gamma$-mode is present at high pressures, high voltage amplitudes, and/or low driving frequencies. In this mode, most ionization is caused by avalanches launched by electrons emitted from the electrode surfaces, which are accelerated and multiplied by collisions inside the sheaths.

\noindent
Electron emission from both conducting and insulating surfaces can be initiated by different species: ions, fast neutrals, metastables, as well as photons. The contributions and the specific yields of these species vary for different gases, electrode materials, electrode surface conditions, and for different operating conditions \cite{Phelps}. Handling all the above species and their processes accurately in simulations is difficult (mostly due to the lack of data). Thus, an {\it effective electron yield}, $\gamma$, defined as the ratio of secondary electron current to the ion current at the surface, is typically used. In the following we also adopt this simplified treatment of the secondary electron emission. This effective yield accounts implicitly for the species other than ions as well. Indications have been published that for given materials and surface conditions $\gamma$ depends on the reduced electric field, $E/n$, in argon dc Townsend \cite{Phelps} and dc glow discharges \cite{Donko_Appgamma}.

\noindent
The electrodes of rf discharges can be made of different materials that can have quite different secondary electron yields. The effect of these different yields (the ``Secondary Electron Asymmetry'' effect [SEAE]) in single-frequency capacitive discharges has recently been analyzed by Lafleur {\it et al.} \cite{Lafleur2013}. It has been found that a significant electrical asymmetry (a dc self-bias up to $\sim$ 20\% of the total driving voltage amplitude) can be generated in case of different (realistic) electron yields at the two electrodes. The observations have been explained by a self-amplifying effect of the larger ion flux at the electrode having the higher secondary yield.

\noindent
These observations raise the question about the coupling of the asymmetries produced by the EAE and the SEAE. We investigate this topic in argon discharges by kinetic particle simulations based on a one-dimensional (1d3v) bounded plasma Particle-in-Cell (PIC) code complemented by Monte Carlo treatment of collision processes (PIC/MCC\cite{Birdsall} method). The cross section sets for electron-neutral and ion-neutral collision processes are taken from Ref.[\onlinecite{Phelps3}]. The discharge is driven by the voltage waveform defined by eq. (\ref{eq:EAEVolt}). The dc self bias, $\eta$, is determined in an iterative manner to ensure that the (positive and negative) charged particle fluxes to either of the two electrodes, averaged over one low frequency period, are equal. At the planar, parallel, and infinite electrodes, electrons are reflected with a probability of 20\%, and we account for the emission of secondary electrons by using secondary yields per incoming ion at the powered and grounded electrodes, $\gamma_{\rm p}$ and $\gamma_{\rm g}$, respectively. We vary $\gamma$ from 0 to 0.4. Note that $\gamma \sim 0.1$ is typical for metal surfaces, $\gamma \sim 0.4$ corresponds to dielectric or semiconductor electrodes. Simulations are performed for $f$ = 13.56 MHz, $p$ = 50 Pa pressure, and $L$ = 2.5 cm electrode gap. We investigate (i) the SEAE without EAE by simulating a single frequency discharge ($\phi_1 \ne 0\, \rm V$, $\phi_2 = 0\, \rm V$, $\gamma_p \ne \gamma_g$), (ii) the EAE without SEAE in a dual-frequency plasma ($\phi_1= \phi_2$ $\gamma_p = \gamma_g$), and (iii) the simultaneous presence of the EAE and SEAE ($\phi_1= \phi_2$, $\gamma_p \ne \gamma_g$). The normalized dc self bias is defined as $\bar{\eta} = \eta/(\phi_1+\phi_2)$.

\noindent
The interpretation of the simulation results is performed on the basis of an analytical model of CCRF discharges.\cite{Heil_b} Here, only its result for the dc self bias, $\eta$, is used in case of identical electrode surface areas:
\begin{equation}
\eta \approx  - \frac{\phi_{\rm max} + \varepsilon \phi_{\rm min}}{1+\varepsilon} \quad \textnormal{with} \quad \varepsilon
  \approx  \frac{\overline{n}_{\rm sp}}{\overline{n}_{\rm sg}}
  \left( \frac{Q_{\rm mg}}{Q_{\rm mp}} \right)^2.
  \label{eq:eta}
\end{equation}

\noindent
Here, $\phi_{\rm max}$ and $\phi_{\rm min}$ is the maximum and the minimum of the driving voltage waveform, $\varepsilon  = \left| \hat{\phi}_{\rm sg}/\hat{\phi}_{\rm sp} \right|$ is the symmetry parameter defined as the ratio of the maximum sheath voltages at both electrodes, $\hat{\phi}_{\rm sp}$ and $\hat{\phi}_{\rm sg}$. $\overline{n}_{\rm sg}$ and $\overline{n}_{\rm sp}$ is the spatially averaged ion density, while $Q_{\rm mg}$ and $Q_{\rm mp}$ is the maximum (uncompensated) charge in the respective sheath.

\begin{figure}[h!]
\begin{center}
\includegraphics[width=0.38\textwidth]{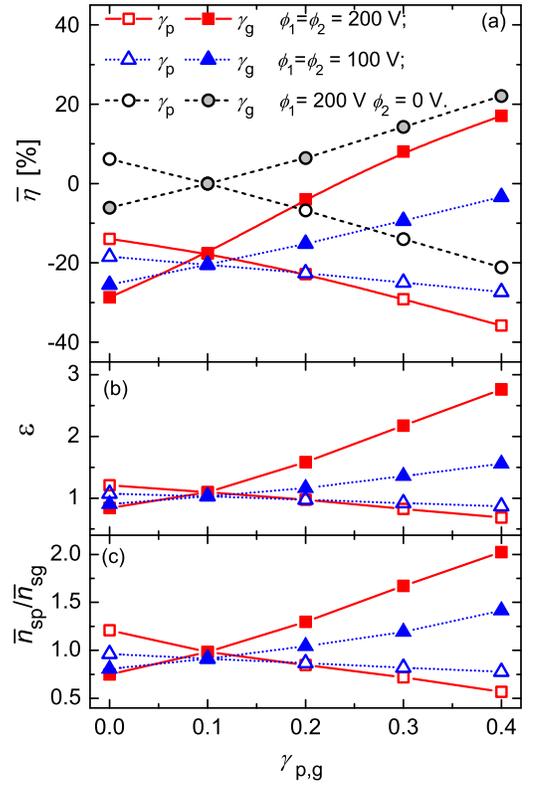}
\caption{PIC/MCC results for the normalized dc self bias, $\bar{\eta}$ (a), symmetry parameter (b) and ${\overline{n}_{\rm sp}}/{\overline{n}_{\rm sg}}$ (c) as a function of the secondary yields ($f$ = 13.56 MHz, $p$ = 50 Pa, $\theta=0^\circ$). Filled symbols: fixed $\gamma_{\rm g}$ = 0.1, open symbols: fixed $\gamma_{\rm p}$ = 0.1.}
\label{fig:bias}
\end{center}
\end{figure}

\noindent
Figure~\ref{fig:bias}a shows $\bar{\eta}$ as a function of the secondary electron yield at one electrode, $\gamma_{\rm p,g}$, while the emission coefficient at the other electrode is kept constant, i.e. $\gamma_{\rm g,p}$ = 0.1. Results for different harmonics' amplitudes are shown for $\theta = 0^\circ$. We have included data for a single frequency discharge with the amplitude $\phi_1$ = 200 V ($\phi_2$ = 0 V), where the self bias is generated due to the SEAE without the presence of the EAE (dashed lines and circles), such as observed by Lafleur {\it et al.} \cite{Lafleur2013}. We find, that no DC self bias is generated for $\gamma_p = \gamma_g$ in single frequency discharges. For $\gamma_{p,g} = 0.4$ and $\gamma_{g,p} = 0.1$ we find $\bar{\eta} \approx \pm$ 20 \%. For $\gamma_p = \gamma_g = 0.1$ and using identical harmonics' amplitudes of 200 V or 100 V, a dc self bias of approximately 20 \% is generated via the EAE only. Thus, under the conditions investigated here, the EAE and the SEAE individually lead to $\bar{\eta} \approx 20 \%$. Combining the EAE and the SEAE, i.e. using $\phi_1= \phi_2$ and $\gamma_p \ne \gamma_g$, allows to enhance or reduce $\bar{\eta}$ with respect to the EAE by tuning $\gamma_{p,g}$. This effect is more pronounced at higher voltage amplitudes. Generally, the dc self bias generated by the EAE and SEAE individually will not simply add up, if both effects are present simultaneously, i.e. their superposition is non-linear.

\begin{figure}[h!]
\begin{center}
\includegraphics[width=0.47\textwidth]{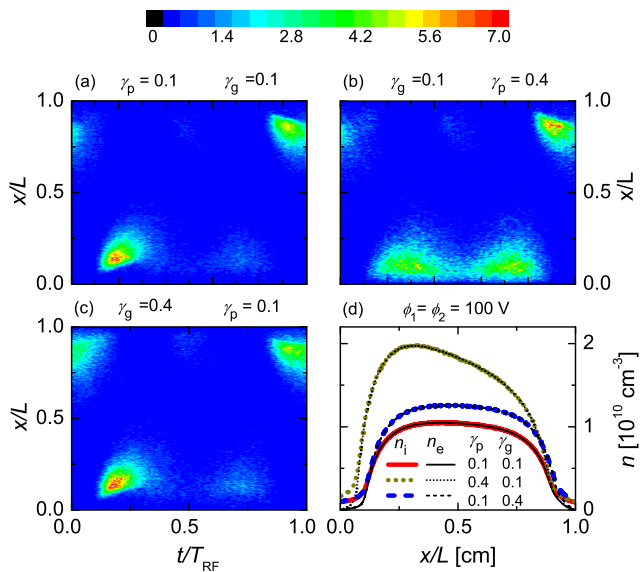}
\caption{Spatio-temporal ionization rate in units of $10^{21}$m$^{-3}$s$^{-1}$ [(a)-(c)] and time averaged ion and electron density (d) for different pairs of secondary yields ($p$ = 50 Pa, $\phi_1 = \phi_2$ = 100 V, $\theta = 0^\circ$). The powered electrode is situated at $x/L=0$, while $x/L=1$ corresponds to the grounded electrode.}
\label{fig:xt}
\end{center}
\end{figure}

\noindent
The coupling of the EAE and the SEAE can be understood based on the model, i.e. eq. (\ref{eq:eta}): Figures~\ref{fig:bias}b and c show the symmetry parameter, $\varepsilon$, and the ratio of the mean ion densities, $\overline{n}_{\rm sp}/\overline{n}_{\rm sg}$, in both sheaths, respectively. According to eq. (\ref{eq:eta}) a dc self bias will be generated, if $\varepsilon \ne 1$ and/or $\phi_{\rm max} \ne -\phi_{\rm min}$. For the EAE excitation (with equal $\gamma$ values at both electrodes), for our conditions (relatively high pressure) $\varepsilon \approx 1$ and the unequal positive/negative extrema of the applied voltage waveform play the dominant role for establishing $\eta$. In a single frequency discharge ($\phi_{\rm max} = -\phi_{\rm min}$) with unequal $\gamma$ values (SEAE) the deviation of the symmetry parameter from one creates the self bias. If both effects (EAE and SEAE) are present, both $\phi_{\rm max} \ne -\phi_{\rm min}$ and $\varepsilon \ne 1$ will contribute to $\eta$. The SEAE, i.e. $\gamma_p \ne \gamma_g$, causes the ionization rate by secondary electrons to be higher at the electrode, where the $\gamma$-coefficient is higher, compared to the other electrode such as shown in Figure~\ref{fig:xt}. For low values of $\gamma$ at both electrodes (Figure~\ref{fig:xt}a) we observe characteristic patterns for an $\alpha$-mode discharge, i.e. ionization at the edge of the expanding sheath dominates. At strongly different $\gamma$ values at both sides of the discharge we observe dominant $\alpha$-ionization at the low-$\gamma$ side, and a dominant $\gamma$-ionization at the high-$\gamma$ side. Thus, e.g., for $\gamma_{\rm p} \textgreater \gamma_{\rm g}$, the ionization by secondary electrons is more effective at the powered electrode (see Figure~\ref{fig:xt}c), which affects the discharge symmetry, i.e. $\varepsilon$ increases due to an increase of $\overline{n}_{\rm sp}/\overline{n}_{\rm sg}$ such as shown in Figures~\ref{fig:bias}b and c. The time averaged density distributions for different pairs of the secondary yields are shown in Figure~\ref{fig:xt}d. At higher voltages the maximum sheath voltage is higher. This further enhances the ionization by secondary electrons at the time of maximum sheath voltage and amplifies the SEAE. $\varepsilon$ is not only determined by the density ratio, but also by $(Q_{\rm mg}/Q_{\rm mp})^2$. For $\theta = 0^\circ$, $(Q_{\rm mg}/Q_{\rm mp})^2 > 1$ due to the charge dynamics \cite{QDyn}. Its value is, however, nearly constant independent of $\gamma_{p,g}$.

\begin{figure}[h!]
\begin{center}
\includegraphics[width=0.35\textwidth]{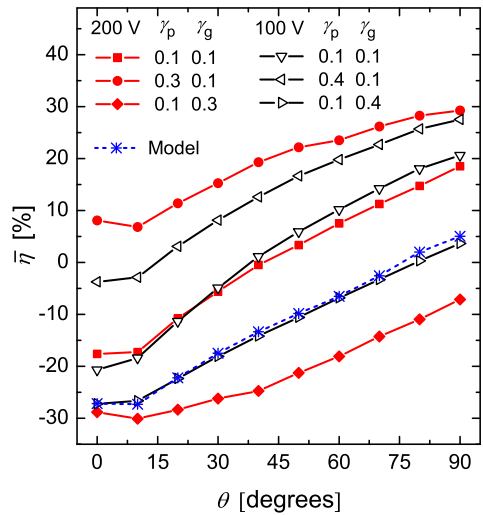}
\caption{$\bar{\eta}$ as a function of $\theta$ for different pairs of $\gamma_{\rm p}$ and $\gamma_{\rm g}$ ($\phi_1 = \phi_2 $ = 100 V, 200 V) resulting from the PIC simulations (solid lines) and from the analytical model via eq. (\ref{eq:eta}) (dashed line) for $\phi_1 = \phi_2$ = 100 V, $\gamma_{\rm p}$ =0.1 and $\gamma_{\rm g}$ = 0.4.}
\label{fig:biasphase}
\end{center}
\end{figure}

\begin{figure}[h!]
\begin{center}
\includegraphics[width=0.35\textwidth]{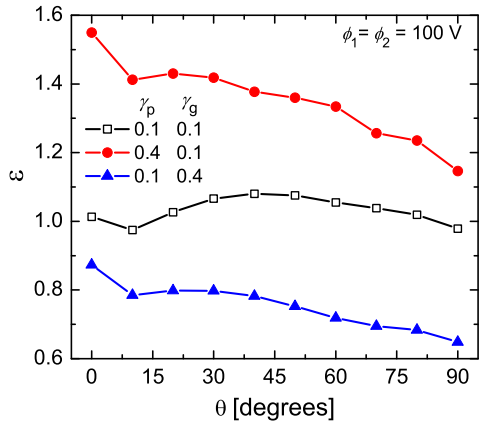}
\caption{$\varepsilon$ as a function of $\theta$ for different pairs of $\gamma_{\rm p}$ and $\gamma_{\rm g}$ ($p$ = 50 Pa, $\phi_1 = \phi_2$ = 100 V).}
\label{fig:eps100}
\end{center}
\end{figure}

\noindent
Next, we illustrate the dependence of $\bar{\eta}$ on $\theta$ (see Figure~\ref{fig:biasphase}). The solid lines show the simulation results, whereas the dashed line corresponds to the model calculation for a selected case using eq.(\ref{eq:eta}). The good agreement between the model and simulation results shows that the model will still be reliable, if both the EAE and the SEAE are present simultaneously. Implementing the SEAE leads to a shift of the control range of $\bar{\eta}$. The shift is not constant for different values of $\theta$ and fixed values of $\gamma_{\rm p,g}$. This shows that the dc self bias generated via the SEAE cannot be simply added to, or subtracted from the self bias generated via the EAE (non-linear coupling). Figure~\ref{fig:eps100} displays the symmetry parameter as a function of $\theta$. We observe that depending on $\theta$, the SEAE changes the symmetry to a different extent, which can be explained as follows: The dc self bias, $\eta = | \left< \phi_{sg} \right> |- | \left< \phi_{sp} \right> | $, corresponds to the difference of the time averaged sheath voltages and changes as a function of $\theta$. This difference strongly affects the ionization by secondary electrons at each electrode. For instance, if $\gamma_p > \gamma_g$ and $\bar{\eta} < 0$ at $\theta = 0^\circ$, the ionization by secondary electrons at the powered electrode will be enhanced compared to $\bar{\eta} > 0$ at $\theta = 90^\circ$. Thus, $\bar{n}_{sp} > \bar{n}_{sg}$ and $\varepsilon > 1$ at $\theta = 0^\circ$ and $\varepsilon$ decreases as a function of $\theta$. Consequently, the maximum sheath voltages at the powered and at the grounded electrode change as a function of $\theta$.

\begin{figure}[h!]
\begin{center}
\includegraphics[width=0.43\textwidth]{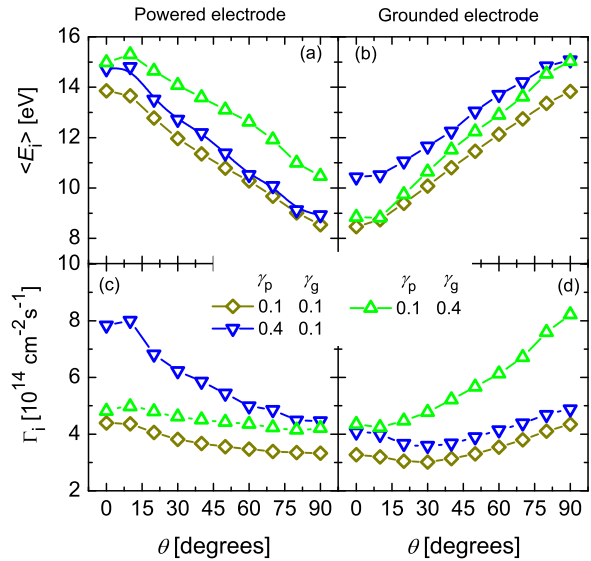}
\caption{$\langle E_i \rangle$ (top) and $\Gamma_i$ (bottom) as a function of $\theta$ for different pairs of $\gamma_{\rm p}$ and $\gamma_{\rm g}$ ($p$ = 50 Pa, $\phi_1 = \phi_2$ = 100 V).}
\label{fig:enfl100}
\end{center}
\end{figure}

\noindent
Finally, we present the dependence of $\langle E_i \rangle$ and $\Gamma_i$ on $\theta$ for different pairs of the secondary yields at both electrodes at $\phi_1 = \phi_2$ = 100 V. Figure~\ref{fig:enfl100} shows, that the control range for the mean ion energy at one electrode can be enlarged by combining the EAE and the SEAE. However, for $\gamma_{\rm g} \neq \gamma_{\rm p}$ the ion flux does no longer remain constant as a function of $\theta$ due to changes of the ionization dynamics induced by changing $\theta$. Although there is no separate control of ion properties under these conditions, this scenario might be ideal for radio frequency sputtering, where typically different materials are used for both electrodes (substrate and target). $\langle E_i \rangle$ could be minimized at the substrate (low ion flux), while it is maximized at the target at a high ion flux.

\noindent
In summary, we have investigated the effect of unequal electron yields of the two electrodes of capacitive radio frequency discharges operated under the conditions of the EAE. We find the electrical generation of the dc self bias via the EAE to be significantly enlarged or suppressed via the SEAE, if two electrodes have noticeably different $\gamma$. The EAE and the SEAE couple non-linearly. At such conditions, the control range of $\langle E_i \rangle$ as a function of $\theta$ can be enlarged, while $\Gamma_i$ does no longer remain constant. Such statements are true for the conditions investigated here, but cannot be generalized without further studies covering a wider range of gas pressures. Experimental investigations of the effect are also required.

\vspace{0.2cm}

\noindent
This work was supported by the Hungarian Scientific Research Fund (OTKA-K-77653, IN-85261, and K-105476).

\appendix

\providecommand{\noopsort}[1]{}\providecommand{\singleletter}[1]{#1}%

\end{document}